\documentstyle{cupconf}
\input epsf

\begin{document}
\title[Turbulence in the WIM]{Probing Interstellar Turbulence in the 
Warm Ionized Medium using Emission Lines}

\author[S. L. Tufte et al.]{Stephen L. Tufte, Ronald
J. Reynolds, \& L. Matthew Haffner}
\affiliation{Department of Astronomy, University of Wisconsin -- 
        Madison \\475 N. Charter St., Madison, WI 53706}
\maketitle

\begin{abstract}
The nature of turbulence in the warm ionized component of the
interstellar medium (WIM) can be investigated using Fabry-Perot
spectroscopy of optical emission lines.  The H$\alpha$ intensity
provides the emission measure (EM) along a line of sight, which is
used in conjunction with the scattering measure, rotation measure, and
dispersion measure to study interstellar turbulence.  Observations at
high spectral resolution ($\sim$ 10 km s$^{-1}$) allow measurements of
the bulk radial velocity structure of the emitting gas and
investigations of thermal and non-thermal (turbulent) broadening
mechanisms through the line widths.  By measuring the widths of the
H$\alpha$ line and an emission line from a heavier atom (e.g.\ the
[S~II] $\lambda$6716 line), one can separate the thermal and
non-thermal contributions to the line width.  Preliminary studies
using this method have shown that the broad range of H$\alpha$ line
widths (typically 15 -- 50 km s$^{-1}$) is mostly due to differences
in the non-thermal component of the width and that along many lines of
sight this component dominates.  The Wisconsin H$\alpha$ Mapper (WHAM)
is in the process of producing a very sensitive kinematic map of the
northern sky in H$\alpha$ at 1$^{\rm o}$ angular resolution and 12 km
s$^{-1}$ spectral resolution.  WHAM is also mapping emission lines
from heavier atoms such as sulfur and nitrogen for selected regions of
the sky.  This data set will provide unique new information concerning
turbulence in the WIM.

\end{abstract}

\firstsection 

\section{Introduction}
The presence of ionized gas in the Galaxy has traditionally been
associated with the bright ionized regions near hot stars called
Str\"{o}mgren spheres or classical H II regions.  We now know that
these classical H II regions contain only about 10\% of the of the
ionized hydrogen in the Galaxy.  The remaining 90\% is in the form of
warm ($\sim$~10$^{4}$~K), low density ($\sim$ 0.1 cm$^{-3}$), fully
ionized regions which fill approximately 20\% of the volume within a 2
kpc thick layer about the Galactic plane.  This warm ionized medium
or WIM (sometimes also referred to as the Diffuse Ionized Gas or DIG),
is now recognized as a major component of the interstellar medium
(e.g. \cite{Kulkarni87}).

The primary sources of information about the WIM are pulsar dispersion
measures and faint optical emission lines, most notably the H$\alpha$
line.  The amount of dispersion in pulsar signals is directly related
to the quantity of free electrons along the line of sight toward the
pulsar, quantified by the dispersion measure DM $\equiv$
$\int_{0}^{L}$ n$_{\rm e}$ $dl$.  The H$\alpha$ emission results from
recombination in ionized regions and is thusly related to the
integrated electron density {\em squared}, quantified by the emission
measure EM $\equiv$ $\int_{0}^{\infty}$ n$_{\rm e}^2$ $dl$.

The physical properties of interest for studies of turbulence in the
WIM are the {\em fluctuations} in the electron density, magnetic
field, and velocity field.  The electron density fluctuations
$\delta$n$_{\rm e}$ are the easiest to probe and are thought to trace
the dynamically more important magnetic field and velocity
fluctuations.  These electron density fluctuations are quantified
through the scattering measure SM.  This quantity is observationally
probed using radio frequency measurements of intensity scintillations
and angular broadening of galactic and extragalactic sources (see
the Cordes paper in this volume for details).

This paper will summarize the potential of the new Wisconsin H-Alpha
Mapper (WHAM) facility for increasing our understanding of
interstellar turbulence in the WIM.  The relevance of the WHAM data to
turbulence studies is three fold: 1) the H$\alpha$ survey provides
measurements of the emission measure EM for the entire northern sky
for correlation studies with turbulent parameters such as the
scattering measure, 2) the WHAM maps allow the study of spatial
fluctuations in the WIM, and 3) the high spectral resolution of the
WHAM spectrometer provides the opportunity to study both the bulk
velocity structure of the WIM through velocity interval maps and
thermal and non-thermal line broadening mechanisms through study of
the spectral line widths.  Each of these topics is explored further
below.

\section{The WHAM Facility}
The recently completed Wisconsin H-Alpha Mapper (WHAM) facility is a
powerful new tool for the study of very faint optical emission lines
from the diffuse interstellar medium (\cite{Reynolds97}). WHAM consists
of a 15-cm aperture, dual-etalon Fabry-Perot spectrometer coupled to a
0.6-m aperture siderostat.  WHAM is located at Kitt Peak, AZ and is
operated remotely from Madison, WI.

The primary purpose of the WHAM facility is to conduct a sensitive
kinematic survey of the northern sky ($\delta$ $>$ $-30^{\rm o}$) in
the H$\alpha$ line with 1$^{\rm o}$ spatial resolution and 12 km
s$^{-1}$ spectral resolution over a 200 km s$^{-1}$ wide spectral
window centered near the local standard of rest velocity.  The WHAM
H$\alpha$ survey will provide the first detailed look at the
distribution and kinematics of the ionized gas in the Galaxy.  This
data set will be compared to existing H I surveys, which will
allow study of the detailed inter-relationship between the neutral and
ionized gas in the interstellar medium.  The data for the survey have
now been collected and the data reduction and calibration are
currently being refined towards producing accurate all-sky maps to
distribute to the astronomical community.  Although WHAM has been used 
primarily for observations of H$\alpha$, the spectrometer can be tuned 
to any wavelength between 4800 $\AA$ and 7200 $\AA$.  In fact, 
observations of [S~II] $\lambda$6716 and [N~II] $\lambda$6584 have
recently begun for selected areas of the sky (\cite{Haffner98}).

The spectra measured by the new WHAM instrument are much like the
previous Wisconsin Fabry-Perot spectrometer, but the new instrument
can measure a spectrum with the same signal-to-noise ratio in 100
times shorter integration time.  This tremendous increase in
sensitivity is realized through the technique of imaging the ring
pattern passed by the Fabry-Perot etalons onto a CCD camera.  The
previous instrument used a single element (PMT) detector and therefore
could only measure one spectral element at a time.  The new method
utilizes the dependence of the wavelength on the off-axis angle to
measure the entire spectrum (approximately 4.4 $\AA$) all at once.

\begin{figure}
  \vspace{0.125in}
  \centerline{\hbox{
      \epsfysize=2.7in \epsfbox{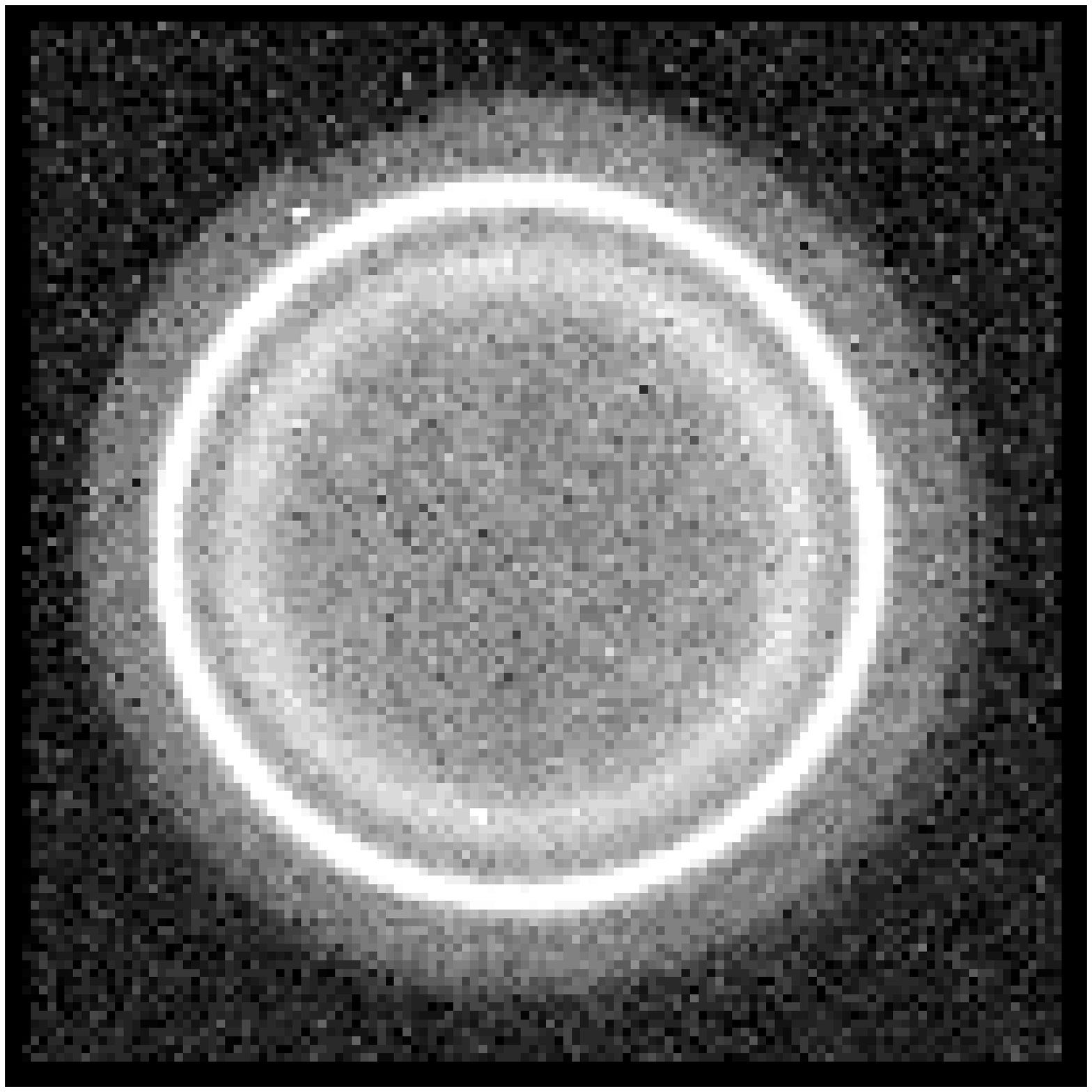}
      \epsfysize=2.7in \epsfbox{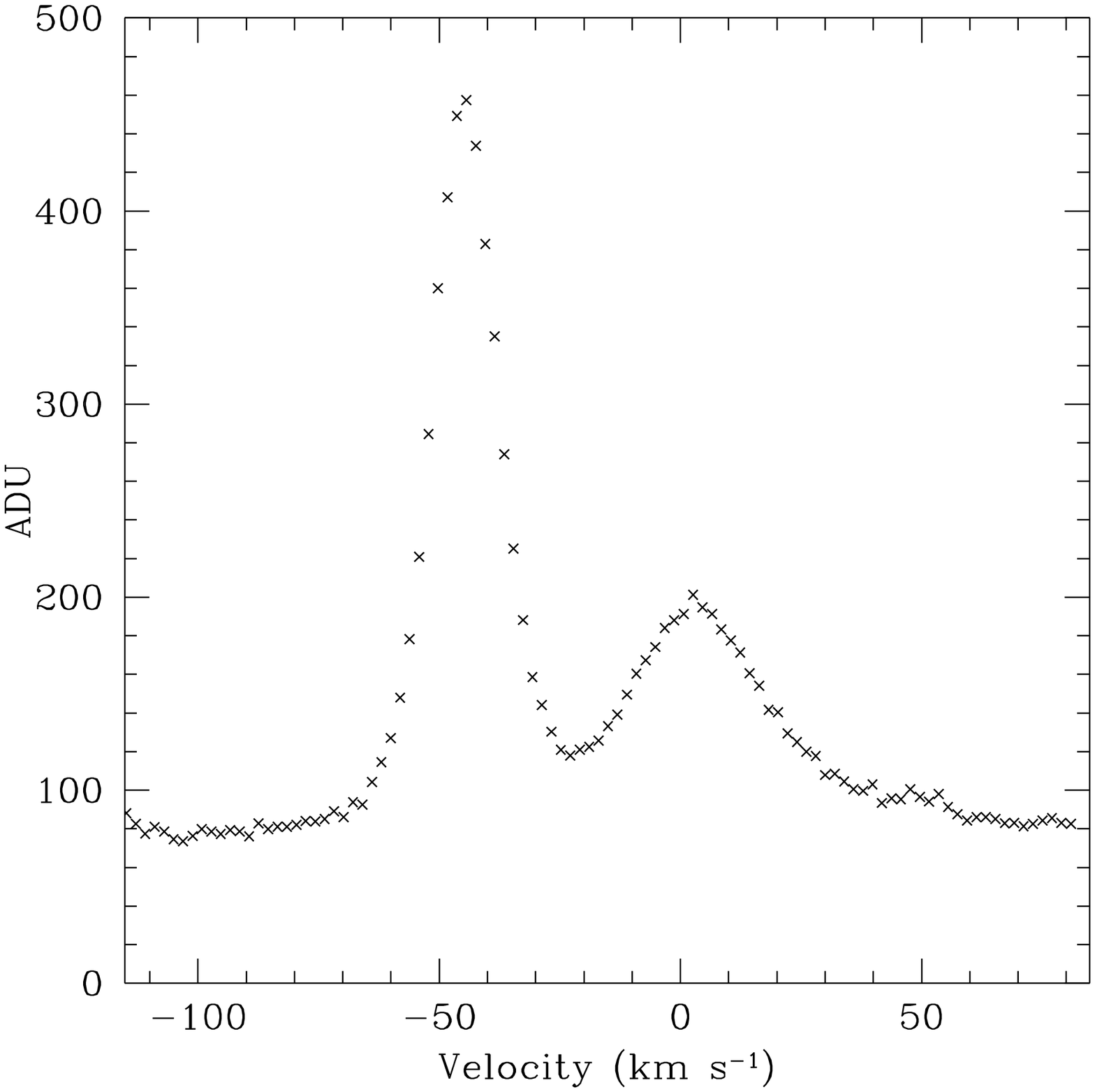}}
    }
  \caption{
    A 30-second Fabry-Perot H$\alpha$ ring image and the resulting spectrum
    toward $l=185^{\rm o}$, $b=-6^{\rm o}$. The diffuse interstellar
    H$\alpha$ emission is centered near 0 km~s$^{-1}$ (LSR) and has an 
    intensity of approximately 8~R.  The narrower feature at 
    $-45$ km~s$^{-1}$ is H$\alpha$ emission from the earth's outer 
    atmosphere (the geocoronal line).}
\label{fig-ringim}
\end{figure}

An example of data collected by the WHAM spectrometer is shown in the
left panel of Figure~\ref{fig-ringim}.  This is the raw CCD image of
the ring pattern passed by the Fabry-Perot etalons acquired during a
30 second exposure toward the direction $l=185^{\rm o}$, $b = -6^{\rm
o}$ (galactic coordinates) using the WHAM system.  The outer ring is
from H$\alpha$ emission from the earth's outer atmosphere (geocoronal
emission) and the inner ring shows H$\alpha$ emission from the diffuse
interstellar medium of our Galaxy.  The right panel of
Figure~\ref{fig-ringim} shows the spectrum derived from this image by
summing pixels in concentric annuli (the red end of the spectrum is
from the center of the ring image and the blue end from the edge.)
The offset between the two components is largely due to the earth's
orbital velocity.  Blending between the geocoronal and Galactic
emission for a given direction can be minimized by carrying out the
observation when the component of the earth's orbital velocity vector
along the line of sight is a maximum.

\section{The Emission Measure}
The H$\alpha$ intensity along a line of sight is directly related to
the emission measure, EM $\equiv$ $\int_{0}^{\infty}$ n$_{\rm e}^2$
$dl$ (neglecting extinction due to dust).  A study by
\cite{Spangler90} investigated the relationship between the emission
measure and the radio scattering disk at 1 GHz (a measure of the
strength of scattering and directly related to the scattering measure
SM).  This study found a correlation between EM and SM, indicating
that the two observations probe the same plasma.  The study also found
significant scatter about the expected relationship suggesting that
the properties of the turbulence varied from sightline to sightline.
These findings were confirmed by \cite{Cordes91} in a study which
compared SM to DM.  The latter study found a clear correlation between
SM and DM which steepened significantly beyond DM $\simeq$ 80 pc
cm$^{-3}$.  These studies make clear that in order to isolate the
behavior of the turbulent properties of the ionized interstellar
medium, it is important to know the amount of ionized gas along the
lines of sight in addition to measures of the fluctuations.  The WHAM
H$\alpha$ survey data will provide a rich resource for use in future
studies of this sort.

Another way in which the EM has been used is to separate the n$_{\rm
e}$ and B components of variations seen in the Faraday rotation measures 
(RM $\propto$ $\int_{0}^{L} n_{e} B_{\parallel}$ ds) toward extragalactic
radio sources.  \cite{Minter96} measured Faraday rotation measures
toward 38 extragalactic sources in a region of the sky for which
H$\alpha$ intensity maps existed (\cite{Reynolds80}).  This allowed
the first measurement of the magnetic field fluctuations in the WIM.
This information combined with measurements of the outer scale and
spectrum of the turbulent fluctuations allowed the calculation of the
energy density in the turbulence.  For a more detailed report on these
findings, see Minter \& Balser (this volume).

\section{Non-thermal Line Broadening in the WIM}
The measured widths of the H$\alpha$ line from the WIM range from 15
to 50 km s$^{-1}$ and result from thermal and non-thermal broadening
mechanisms.  By measuring spectra of both H$\alpha$ and emission from
a heavier atom such as the [S~II] $\lambda$6716 line, one can separate
the thermal and non-thermal contributions to the line width.
\cite{Reynolds85} carried out such a study of the warm ionized medium
using data from 21 directions observed with the PMT based
spectrometer.  This pilot study showed that the temperature, while not
tightly constrained, appears to be around 8000 K and that the
non-thermal velocities have a broad range v$_{nt}$ $\simeq$ 10 - 30 km
s$^{-1}$.

The WHAM spectrometer has already mapped a large region of the sky
($\sim$ 2000 spectra) in the [S~II] $\lambda$6716 line with higher
signal-to-noise than was possible with the previous spectrometer.
Using these new data together with the H$\alpha$ survey data, we will be
able to form {\em maps} of the non-thermal velocity widths, allowing
detailed study of the spatial and kinematic behavior of this quantity.
Some of the non-thermal velocity width is due to smearing caused by
differential Galactic rotation, but at high latitudes and at certain
galactic longitudes, this effect is minimized allowing the study of
the ``turbulent'' broadening component.  It will be interesting to see
how well v$_{nt}$ correlates with the scattering measure, for example.

\section{Turbulent Heating in the WIM?}
\begin{figure}[tb]
  \centerline{ 
  \epsfysize = 4.0in 
  \epsffile{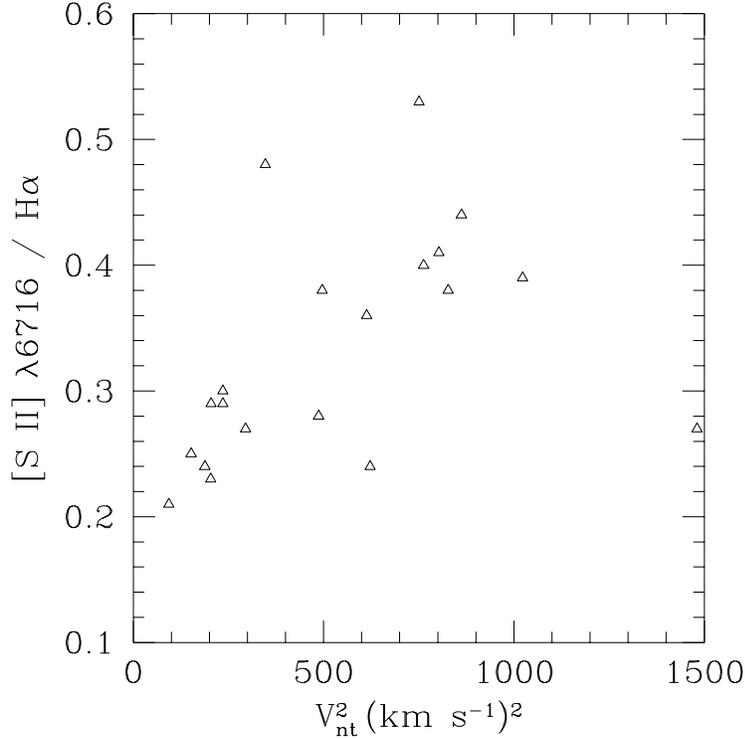} }
    \caption{[S~II] $\lambda$6716 / H$\alpha$ vs. v$_{nt}^2$.  The
    correlation seen here may indicate that regions within the WIM
    that have stronger turbulence are also higher in temperature.}
\label{fig-sii}
\end{figure}

Using the results of \cite{Minter96}, \cite{Minter97a} have shown that
the dissipation of turbulent energy in the WIM is a potentially
important additional source of heating.  \cite{Minter97b} were able to
reproduce the peculiar line ratios seen in the WIM using a model
incorporating the dissipation of turbulence as an additional heating
source.  As a further indication that this may in fact be occurring,
note that in data taken from \cite{Reynolds85}, the [S~II]
$\lambda$6716 / H$\alpha$ ratio has been found to correlate with the
non-thermal width v$_{nt}$ discussed above (Fig.~\ref{fig-sii}; also
see Fig. 2 in the Minter \& Balser paper in this volume).  Since the
[S~II] $\lambda$6716 line is excited by thermal electrons, it will be
brighter in regions with higher temperature.  The correlation with
v$_{nt}$ suggests that regions with strong turbulence have higher
temperature, corroborating the suggestion that dissipation of
turbulence is a source of heating.  It should be noted, however, that
variations in the brightness of the [S~II] $\lambda$6716 line can also
occur due to ionization effects (i.e. variations in the fraction of
sulfur that is in the singly ionized state.)  The relationship between
[S~II] / H$\alpha$ and v$_{nt}$ will be further investigated using the
much more extensive and higher signal-to-noise WHAM data.  Also, the
[N~II] $\lambda$6584 / H$\alpha$ ratio can simultaneously be measured
as a control on whether the [S~II] $\lambda$6716 / H$\alpha$ ratio is
changing due to temperature variations, or an ionization effect.  We
have, in fact, already mapped approximately 1000 square degrees of the
sky in all three of these lines (\cite{Haffner98}).

\section{Large-scale Spatial and Velocity Structure}
\begin{figure}[p]
  \centerline{
    \epsfysize = 7.0in
    \epsffile{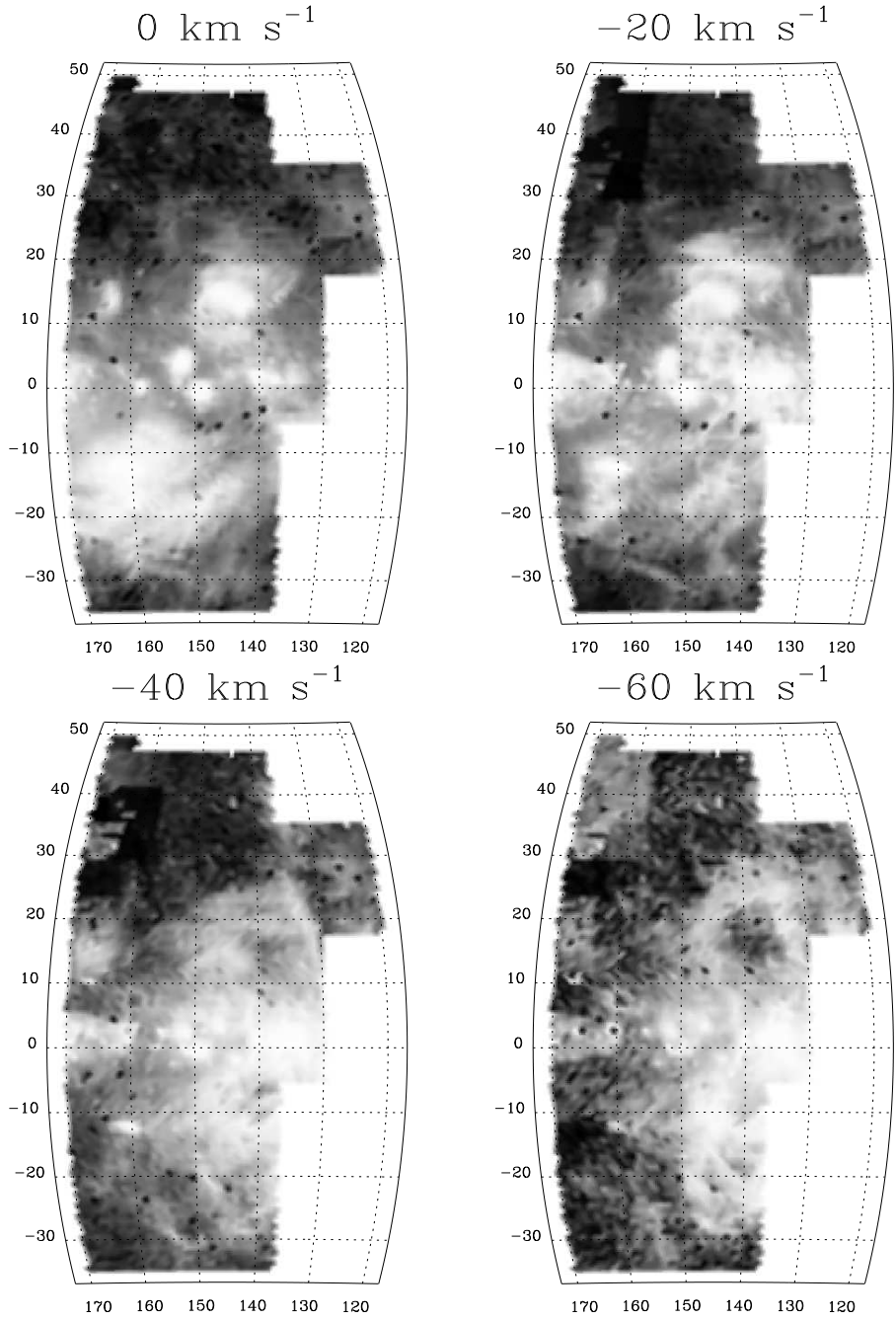}
    }
  \caption{H$\alpha$ toward the Perseus Arm showing rich spatial and 
    velocity structure.  Each map covers a 12 km s$^{-1}$ velocity 
    interval.  Coordinates are in $l$ and $b$.  The small black dots are
    pixels contaminated by bright stars.}
\label{fig-velmap}
\end{figure}

The ultimate energy source for turbulence in the interstellar medium
is believed to be a combination of supernova explosions, winds from
massive stars, and differential Galactic rotation.  In all cases the
energy is input on large spatial scales and must in some way cascade
to smaller scales until it ultimately is thermalized and appears as
heat.  Fluctuations on the intermediate spatial scales can be
investigated with the WHAM survey data in two ways.  First, the
spatial structure revealed in the survey data will reveal how clumpy
the WIM is on scales larger than 1$^{\rm o}$.  Second, the velocity
structure revealed by the velocity interval maps measures the
kinematics of the emission regions.  This is illustrated in
Figure~\ref{fig-velmap}, which shows WHAM velocity interval maps from
a region of the sky toward the Perseus Arm.  The four selected
velocity interval maps are 12 km s$^{-1}$ wide and are centered at
velocities ranging from 0 km s$^{-1}$ to --60 km s$^{-1}$.  Near the
local standard of rest (0 km s$^{-1}$) the maps reveal the structure
and kinematics of local gas within a few hundred parsecs of the Sun.
Two large, low surface brightness H~II regions associated with
high-latitude O stars, $\alpha$ Cam at $l$ = 144$^{\rm o}$, $b$ =
+14$^{\rm o}$ and $\xi$ Per at $l$ = 160$^{\rm o}$, $b$ = -13$^{\rm
o}$, dominate large portions of this map.  At higher negative
velocities (-40 km s$^{-1}$ to -60 km s$^{-1}$) the region is
dominated by the diffuse ionized hydrogen in the Perseus spiral arm
located 2 -- 3 kpc from the Sun.  Much of this gas appears to be
associated with enormous arch-like features extending 1300 pc
(30$^{\rm o}$) away from the Galactic midplane.  Rich kinematic
structure within these features is revealed by a comparison between
the -40 km s$^{-1}$ map and the -60 km s$^{-1}$ map.

\begin{acknowledgments}
This work has been supported by the National Science Foundation 
through grants AST9619424 and AST9122701.
\end{acknowledgments}

\end{document}